\documentclass[10pt,a4paper]{article}

\usepackage{amsmath}	
\usepackage{amssymb}	
\usepackage{amsfonts}					
\usepackage{amscd}						
\usepackage{bm}							
\usepackage{mathrsfs}					
\usepackage{graphicx}	
\usepackage{cite}		
\usepackage{color} 
\usepackage[tableposition=bottom,figureposition=bottom]{caption}
\usepackage{enumerate}	
\usepackage{subcaption}
\usepackage{booktabs} 
\usepackage{multirow}
\usepackage{indentfirst}              	
\usepackage{bmpsize}
\usepackage[margin=0.9in]{geometry}

\DeclareMathOperator*{\argmax}{\arg\!\max}

\usepackage{array}
\newcolumntype{P}[1]{>{\centering\arraybackslash}p{#1}}
\newcolumntype{M}[1]{>{\centering\arraybackslash}m{#1}}

\begin{document}

\title{Detection of multiple and overlapping bidirectional communities within large, directed and weighted networks of neurons}
\date{}
\author{
Umberto Esposito \hspace{2.2cm} \footnotesize{\texttt{acp12ue@sheffield.ac.uk}}\\ \footnotesize{University of Sheffield, Department of Computer Science, Sheffield, UK} \vspace{0.2cm} \\
Eleni Vasilaki \hspace{2.5cm} \footnotesize{\texttt{e.vasilaki@sheffield.ac.uk}}\\ \footnotesize{University of Sheffield, Department of Computer Science, Sheffield, UK}\\ \footnotesize{University of Sheffield, INSIGNEO Institute for in Silico Medicine, Sheffield, UK}\\ \footnotesize{University of Antwerp, Theoretical Neurobiology and Neuroengineering Laboratory, Department of Biomedical Sciences, Belgium}}

\maketitle

\begin{abstract}
With the recent explosion of publicly available biological data, the analysis of networks has gained significant interest. In particular, recent promising results in Neuroscience show that the way neurons and areas of the brain are connected to each other plays a fundamental role in cognitive functions and behaviour. Revealing pattern and structures within such an intricate volume of connections is a hard problem that has its roots in Graph and Network Theory. Since many real world situations can be modelled through networks, structures detection algorithms find application in almost every field of Science. These are NP-complete problems; therefore the generally used approach is through heuristic algorithms. Here, we formulate the problem of finding structures in networks of neurons in terms of a community detection problem. We introduce a definition of community and we construct a statistics-based heuristic algorithm for directed and weighted networks aiming at identifying overlapping bidirectional communities in large networks. We carry out a systematic analysis of the algorithm's performance, showing excellent results over a wide range of parameters (successful detection percentages almost $100\%$ all the time). Also, we show results on the computational time needed and we suggest future directions on how to improve computational performance.
\end{abstract}

\section*{Introduction}

In recent years the study of the wiring diagram of the brain has raised enormous attention in Neuroscience \cite{Lichtman2008a,Seung2009,VanEssen2012}. Several experimental works reported significant excess of particular connectivity motifs in different areas of the brain \cite{Song2005,Silberberg:2007,Lefort2009,Perin2011}, suggesting that connectivity is generally not random \cite{Sporns2011}. Moreover, different motifs seem to correlate with different synaptic properties \cite{Wang2006,Pignatelli2009}, which in turn are related to signal transmission, underlying learning mechanisms and eventually cognitive functions and behaviour \cite{Lichtman2008,Bullmore2009,Bressler2010}. It is largely believed that a complete map of the connections between neurons, the so-called connectome \cite{Sporns2005}, could provide an unprecedented and extremely powerful knowledge, with great benefits, for instance, in diseases treatment \cite{Zhou2012,VanEssen2012a,Wang2013}.

It is therefore essential to reveal the structural and functional properties of brain networks. To achieve this, principles and tools from Graph and Network Theory have been widely applied to brain networks \cite{He2010,Sporns2011a,Sporns2013} with promising results \cite{Bassett2009,Guye2010}. Several studies have demonstrated that many real world processes can be modelled in terms of complex networks \cite{Albert2002,Barabasi2004,Green2005,Newman2010}, making the study of networks' topology and properties a topic of major interest within the entire scientific community.

Of particular relevance for brain networks is the problem of structures' detection: sub-regions of networks whose connectivity has been significantly shaped by an underlying learning process. Typical Graph Theory problems dealing with structure searching, for instance sub-graph isomorphism and clique identification \cite{BondyMurty2008,Diestel2010}, are proven to be either NP-complete or NP-hard \cite{Cook1971,Papadimitriou1977,Garey1990,Bomze1999,Wegener2005}. Extensive search is therefore impracticable and feasible approaches are based on heuristic search or on algorithms looking for sub-optimal solutions. Even with these approaches, the computational complexity grows very quickly and explodes for just few thousands of nodes, hence making impossible to perform an effective and accurate search on large networks within a relatively small time scale. The purpose of this work is to contribute in this direction by means of a heuristic algorithm designed to identify a particular class of such structures.

Besides the computational limitations, networks of neurons are arguably the most challenging type of graph to deal with, as they are instances of directed and weighted graphs with continuous weights. Most studied problems in Network Theory are based on undirected \cite{Fortunato2010} networks, with some of them focussing on directed un-weighted \cite{Malliaros2013} (or binary) graphs. In most of the cases generalisation to directed and weighted graphs is not always trivial. Moreover, in general, there is no limitation on the number of structures that can be formed within a network of neurons, nor on their shapes and overlaps. This leads to a very generic problem that needs to be narrowed to design an effective searching algorithm.

On the other side, we show that having a network of neurons and structures that arise from learning allows us to make considerations and hypotheses that greatly simplify the searching task, ultimately framing it within the domain of community detection in Network Theory \cite{Girvan2002, Newman2004}. This field has received constantly increasing attention due to the fact that community structures are often present in many types of networks and through their study the understanding of the network itself can be greatly improved \cite{Porter2009}. However, despite huge efforts of a large interdisciplinary community of scientists, the problem is not yet satisfactorily solved.

Most of the existing algorithms for community structure use techniques like hierarchical clustering \cite{Girvan2002, Newman2004a}, modularity optimisation \cite{Danon2005,Newman2006a,Ovelgonne2012} (which is also a NP-complete problem \cite{Brandes2006}), spectral searching \cite{Newman2013} and statistical inference \cite{Rosvall2007,Ball2011}. These methods are usually not designed for directed and weighted networks and also they do not consider overlapping communities. Furthermore, each class has its own limitations. For instance, modularity optimisation, which is the most widely used method, is known to have resolution problems \cite{Fortunato2007}, and spectral analysis is much more complex for directed graphs as it is characterized by asymmetrical matrices. Developing methods of community detection for directed graphs is a hard task. The most important class of algorithms for the complete problem, i.e. detection of overlapping communities in directed and weighted graphs, is the clique percolation method \cite{Palla2005,Karrer2014}. However, since it does not look for actual communities but just for regions containing many cliques, it fails in several scenarios and its success also depends on the quantity of cliques that are present in the network \cite{Fortunato2010}. Community detection also suffers from the lack of a unique definition: how to identify a community generally varies depending on the problem and on the algorithm, and often a community is just the final outcome of the algorithm itself \cite{Fortunato2010}(\textit{a posteriori} approach).

Here we start by giving a general definition of community (\textit{a priori} approach) and we show how this represents a great advantage as such a definition can be used as a guidance for building the algorithm. Our method, which aims at detecting multiple and overlapping bidirectional communities in directed and weighted networks of neurons, is based on a statistical analysis of connections and it is a mixture of different techniques. At the basis of the algorithm there is the notion of symmetry measure introduced by Esposito et al.\cite{Esposito2014} as an indicator of the global symmetry of a network's connectivity. Below, we introduce a local version of this measure, which, together with the community definition, allows us to develop a peculiar searching technique, a mixture between top-down and bottom-up approaches that does not require looking at single connections to identify communities. This first part already provides very good results and in a very short time, but is able to detect only the non-overlapping parts of communities . Following this, we implement a neuron by neuron evaluation, that we call friendship algorithm, where we restore the detailed information about which pairs of neuron are connected to each other. This greatly increases the total computational time but it also improves the accuracy on the final outcome and allows detecting overlapping regions as part of more than one community.

\section*{Methods}

Consider a directed and weighted network of $N$ nodes that are all-to-all connected, with connectivity matrix $W$. Without losing generality, we allow single connections $w_{ij}$ to vary in $\left[ 0, 1\right]$, where $w_{ij}$ represents the strength of the connection from node $j$ to node $i$. We do not consider self-interactions, i.e. $w_{ii} = 0$ $\forall i = 1 \dots N$.

\subsection*{Preliminary assumptions}
We assume that the network described by $W$ is the result of some learning process that affects only (unknown) parts of the network, significantly shaping these connections away from their initial configuration. Hence:
\begin{itemize}
\item Prior learning, there is no way to differentiate the neurons that are going to be affected by the process from the rest of the network. We therefore assume that before learning all the connections in the network are randomly drawn from the same distribution.
\item Connections between any pair of neurons are subject to the same learning process and therefore evolve in a similar way, which constraints structures to have the particular shape of a blob, rather than, for example, of a filament or a ring.
\end{itemize}

As a result, structures appear as regular bumps that stand out of the global randomness of the network's connectivity. In addition, we take into account that learning can occur with an efficacy $\varphi\leq 100\%$ (some connections may be faulty and not evolve) and that it can be slower for some neurons and faster for others. This makes the final blobs' connectivity far from being a perfect and regular structure: locally, some connections may not display any feature of the learning process, but the majority of the connections in the structure does, which preserves the global property of forming a bump in the network's connectivity.

In what follows, we adopt and generalise the terminology from Network Theory and refer to these structures as communities.

\subsection*{Definition of community and bidirectional community}
For unweighted graphs, a community is generally a region where the edge concentration inside is higher than outside \cite{Fortunato2010}. In this case, the community detection problem for a complete network has the network itself as the only, trivial solution. When searching for communities in continuous weighted networks, it is essential to specify with respect to which property of the connectivity we are investigating the community structure: by changing the feature of the connectivity we look at, a different community structure can emerge. We can therefore phrase the concept of community in terms of an over-expression, in this case, of some property related to the connectivity. Thus, a complete weighted graph, differently from an unweighted one, can present solutions different from the trivial ones and in principle it can offer the same variety in the community structure as a sparse unweighted network.

In our case, the feature we investigate is bidirectionality. Two neurons $i,j$ form a bidirectional pair when both connections have a similar strength, $w_{ij} \simeq w_{ji}$, resulting in information flowing nearly equally in both directions. Guided by experimental results showing excess of bidirectional connections in some regions of the brain \cite{Song2005,Wang2006}, we assume that learning strengthens all the connections involved, thus acting as a Hebbian-like process \cite{Vasilaki2009,Clopath2010,Richmond2011,Vasilaki2012,Vasilaki2014,Esposito2015}. This leads to the formation of what we call bidirectional communities within the network: subsets of neurons that show an over-expression of bidirectional connections among them, when compared with the rest of the network. Since connections are continuous variables, there is no clear way to discriminate a pair that is bidirectional from a pair that is not without using the threshold concept. In the following, we will describe how to fix a bidirectionality threshold, essential for the algorithm implementation.

\paragraph*{Local estimator of bidirectionality.} Esposito et al. 2014 \cite{Esposito2014} introduced a measure of network's connectivity, ranging from $0$ to $1$, that for fully connected networks reduces to the following:
\begin{equation}
\label{sym_meas}
s = 1 -  \frac{2}{N \left( N-1 \right)} \sum_{i=1}^{N} \sum_{j=i+1}^{N} \frac{|w_{ij}-w_{ji}|}{w_{ij}+w_{ji}} \,.
\end{equation}
The extreme values $s = 1$ and $s = 0$ respectively correspond to completely symmetric networks, for which $w_{ij} = w_{ji}$ $\forall \, i,j = 1 \dots N$, and to completely asymmetric networks, for which $w_{ij} = 0$, $w_{ji} \neq 0$ $\forall \, i,j = 1 \dots N$ with $i>j$. In between these extremes, there is a continuum of values with smooth transitions between bidirectional, random and unidirectional networks. Through a statistical analysis of this symmetry measure on random networks, it is possible to identify a \textit{bidirectionality threshold} $s_{B}$, which depends on the distribution of connections, separating bidirectional networks from non-bidirectional ones \cite{Esposito2014}. 

This is, however, a global indicator that cannot capture a deeper organisation at a sub-networks level, nor it can be directly used to find communities, as it would require an extensive search. However, it can be used \textit{i)} to validate community candidates after a successful searching and \textit{ii)} to construct a local estimator encoding for the bidirectionality feature. Indeed, the symmetry measure is a global average of a local pairwise quantity, the relative strength of a pair of connections, defined as:
\begin{equation}
\label{conn_pairs}
Z_{ij} = \frac{|w_{ij}-w_{ji}|}{w_{ij}+w_{ji}}
\end{equation}
$Z$ is a continuous variable ranging from $0$ to $1$ that covers all the possible states in which a connection pair can be found. In particular, bidirectionality is expressed by $Z \to 0$. Similarly to $s$, we can map this continuum into a discrete two-state space, corresponding to randomness and bidirectionality, by fixing a \textit{local bidirectionality threshold} $Z_{B}$ on the connection pair. This can be done by simply translating $s_{B}$ into the corresponding value of $Z$ by using the definition of $s$ itself:
\begin{equation}
\label{bid_thresh}
Z_{B} = 1 - s_{B}
\end{equation}
This follows from the consideration that a network with all equal values of $Z_{ij} = \bar{Z}$, for which $\bar{s} = 1 - \bar{Z}$, must show the same property, for instance bidirectionality, both locally in each pair and globally.

Thus, a bidirectional community of neurons is a set of neurons within which the majority of all possible connection pairs satisfy the relation $Z \geq Z_{B}$, i.e. they are bidirectional.

\paragraph*{Over-density indicator.} The loose concept of majority reflects the over-density property and it can be mathematically formalised by setting a \textit{community threshold} $\vartheta_{\mathcal{C}} \leq 100\%$: a set of neurons is a bidirectional community when, for each neuron within it, at least $\vartheta_{\mathcal{C}}$ of the available connections with other the neurons in the set is bidirectional. This threshold is clearly related to the learning efficacy $\varphi$. For all-to-all connected networks, like the ones we are considering here, we can give the following formal definition:

\textbf{Definition of Community} Be $\mathcal{C}$ a set of neurons, $i \in \mathcal{C}$ a neuron and $\mathcal{S}_\mathcal{C}^{i}:= \left\lbrace j \in \mathcal{C} : \, Z_{ij} \geq Z_{B} \right\rbrace$ the set in $\mathcal{C}$ of all and only the neurons that form a bidirectional pair with neuron $i$. Then $\mathcal{C}$ is a community with respect to the property of bidirectionality if and only if each neuron in $\mathcal{C}$ forms within $\mathcal{C}$ itself at least a number of bidirectional connections equal to a fraction $\vartheta_{\mathcal{C}}$ of the total available connections in $\mathcal{C}$. In formal terms, this can be expressed with the following set of equations:
\begin{equation}
\label{comm_def}
|\mathcal{S}_\mathcal{C}^{i}| \geq \vartheta_{\mathcal{C}} \left( |\mathcal{C}|-1 \right)\quad \forall i \in \mathcal{C} \,,
\end{equation}
where $|\cdot|$ represents the cardinality of a set. Hence, $|\mathcal{S}_\mathcal{C}^{i}|$ is the number of connections between neuron $i$ and the other neurons in $\mathcal{C}$ that are bidirectional, and $|\mathcal{C}|-1$ represents the number of neurons in the community available for forming a bidirectional pair. The maximum value $\vartheta_{\mathcal{C}} = 100 \%$ corresponds to the specific case of a clique, the final result of a perfectly efficient learning $\varphi = 100 \%$. equation (\ref{comm_def}) clearly captures the main difficulty of the community detection problem: we want to find a set of neurons $\mathcal{C}$ whose definition relies on the sets $\left\lbrace \mathcal{S}_\mathcal{C}^{i} \right\rbrace $, which in turn are defined in terms of $\mathcal{C}$ itself and are unknown, with the set $\left\lbrace |\mathcal{S}_\mathcal{C}^{i}| \right\rbrace $ also being unknown.

\begin{table}[h!t]
\begin{center}
\begin{tabular}{M{1.5cm}|M{13cm}}
\toprule
Symbol		&Description\\
\midrule
$N$									&Size of the network\\
$w_{ij}$							&Strength of the single connection from neuron $j$ to neuron $i$\\
$Z_{ij}$							&Relative strength of the connection pair between neurons $i$ and $j$\\
$Z_{B}$								&Bidirectionality threshold for connection pairs\\
$n^{i}$								&Number of bidirectional pairs formed by neuron $i$ in the entire network\\
$\mathcal{P}$						&Bidirectional pool\\
$N_{\mathcal{P}}$					&Size of the bidirectional pool\\
$n^{min}_{\mathcal{P}}$				&Minimum number of bidirectional pairs to be part of the pool\\
$n^{i}_{\mathcal{P}}$				&Number of bidirectional pairs formed by neuron $i$ within the pool\\
$\mathcal{C}$						&Bidirectional community\\
$n^{i}_{\mathcal{C}}$				&Number of bidirectional pairs formed by neuron $i$ within the community\\
$\vartheta_{\mathcal{C}}$			&Threshold for belonging to a community\\
$\mathcal{C}_{i}^{max}$				&Largest possible community that neuron $i$ can form in the pool\\
$\tilde{\mathcal{B}}$				&Bidirectional candidate blob\\
$N_{\tilde{\mathcal{B}}}$			&Size of the bidirectional candidate blob\\
$N^{max}_{\tilde{\mathcal{B}}}$		&Size of the largest community the bidirectional candidate blob can be part of\\
$n^{i}_{\tilde{\mathcal{B}}}$		&Number of bidirectional pairs formed by neuron $i$ within the candidate blob\\
$n^{min}_{\tilde{\mathcal{B}}}$		&Minimum number of bidirectional pairs that each neuron in the candidate blob needs to form within it\\
$\mathcal{B}$						&Bidirectional blob\\
$n^{i}_{\mathcal{B}}$				&Number of bidirectional pairs formed by neuron $i$ within the blob\\
$\tilde{\mathcal{C}}$				&Bidirectional candidate community\\
$N_{\tilde{\mathcal{C}}}$			&Size of the bidirectional candidate community\\
$n^{min}_{\tilde{\mathcal{C}}}$		&Minimum number of bidirectional pairs that a candidate neuron needs to form with the candidate community\\
$n^{i}_{\tilde{\mathcal{C}}}$		&Number of bidirectional pairs formed by neuron $i$ with the members of the current candidate community\\
$\vartheta_{noise}$					&Threshold for noisy communities\\
$\vartheta_{\omega}$				&Threshold for communities merger\\
\bottomrule
\end{tabular}
\end{center}
\begin{small}
\textbf{\refstepcounter{table}\label{tab:symbols}Table 4.\arabic{table}}{ List of the symbols used for the algorithm description and their meaning.}
\end{small}
\end{table}

\subsection*{Algorithm description}
The algorithm we describe below aims at identifying multiple and overlapping bidirectional communities, as defined in equation (\ref{comm_def}), within large networks of neurons. This is achieved by a popularity ranking (\textit{Step 1} below) followed by two different techniques that are applied in sequence (\textit{Step 2} and \textit{Step 3}). If implemented alone, each of them already offers good results, but the combination refines the search and in some cases it also makes it faster.

\begin{figure}[h!t]
\begin{center}
\includegraphics[width=0.95\linewidth]{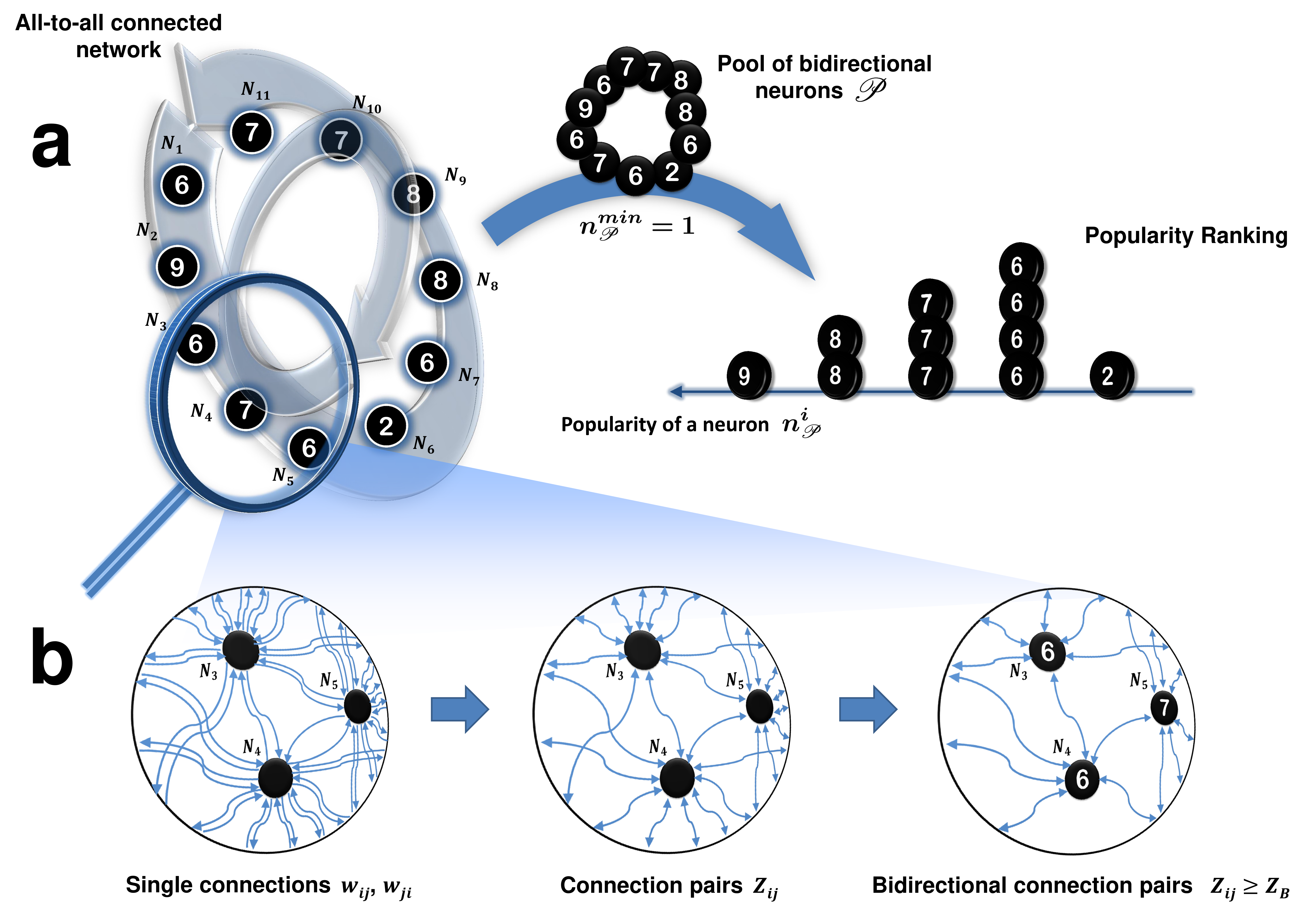}
\end{center}
\caption{{\bf Algorithm Step 1: Neurons popularity ranking.} \textbf{A} All-to-all toy network of $N = 11$ neurons labelled from $N_{1}$ to $N_{11}$ (\textit{left}). Each neuron $i$ is associated with an integer representing the number of bidirectional connections made by $i$ in the entire network, $n^{i}$. Neurons meeting the threshold $n^{min}_{\mathcal{P}} = 1$ are entered in a bidirectional pool $\mathcal{P}$ (\textit{middle}), and here they are sorted in a popularity ranking according to the number of bidirectional connections made within the pool, $n^{i}_{\mathcal{P}}$ (\textit{right}). In this example, network and pool coincide. \textbf{B} Zoom on a portion of the network highlighting the procedure to obtain $\left\lbrace n^{i} \right\rbrace$: The initial directed network $\left\lbrace w_{ij}, w_{ji} \right\rbrace$ (\textit{left}) is mapped into an undirected network of single connection pairs $\left\lbrace Z_{ij} \right\rbrace$ (\textit{middle}). $n^{i}$ counts how many of these pairs fall in the bidirectional domain (\textit{right}). This quantity is used as initial criterion to enter the neurons in the pool $\mathcal{P}$ (see text).}
\label{fig:Algorithm_step1}
\end{figure}

\begin{figure}[h!t]
\begin{center}
\includegraphics[width=0.85\linewidth]{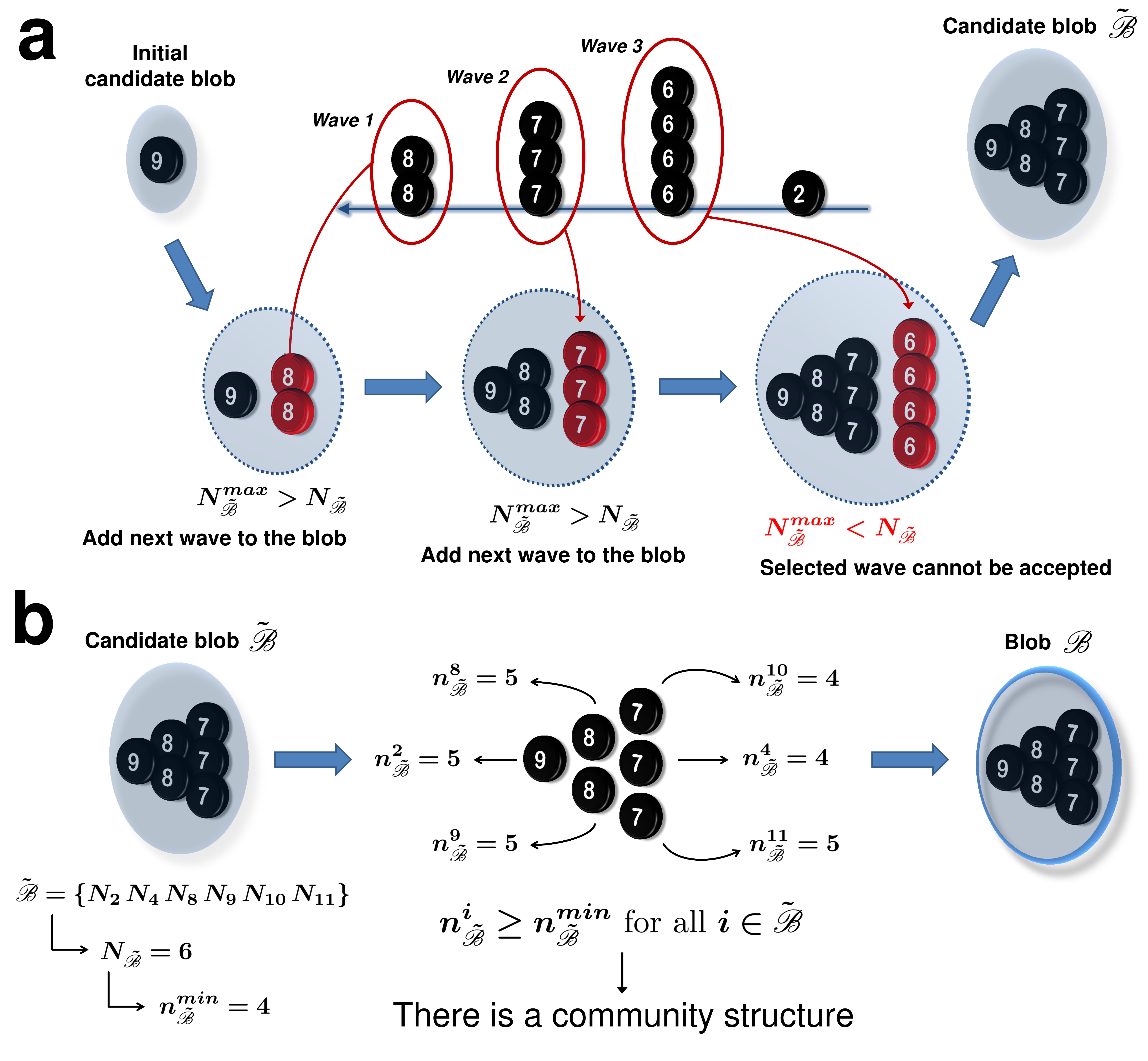}
\end{center}
\caption{{ \bf Algorithm Step 2: Blob search.} \textbf{A} \textit{Part 1}: Detecting a candidate blob $\tilde{\mathcal{B}}$. The research starts with the highest ranked neuron as the only one in $\tilde{\mathcal{B}}$ (\textit{top left}). The other neurons left in the ranking are divided into waves depending on $n^{i}_{\mathcal{P}}$ and are progressively added to the candidate blob until the condition $N^{max}_{\tilde{\mathcal{B}}} > N_{\tilde{\mathcal{B}}}$ is violated (\textit{middle}), see explanation in the text. If $N^{max}_{\tilde{\mathcal{B}}} < N_{\tilde{\mathcal{B}}}$, as in this example, then $\tilde{\mathcal{B}}$ is the set of neurons found before adding the current wave (\textit{top right}). \textbf{B} \textit{Part 2}: Candidate blob validation. The full community definition is restored within $\tilde{\mathcal{B}}$, giving the minimum number of bidirectional pairs that each neuron of the blob needs to form within the blob itself, $n^{min}_{\tilde{\mathcal{B}}}$ (\textit{left}). In this example all the neurons in the candidate blob meet this requirement (\textit{middle}), as they are all-to-all bidirectionally connected except for the pair $N_{4}$, $N_{10}$. The candidate blob satisfies then the complete community definition and gets the status of blob $\mathcal{B}$ (\textit{right}).}
\label{fig:Algorithm_step2}
\end{figure}

\begin{figure}[h!t]
\begin{center}
\includegraphics[width=0.85\linewidth]{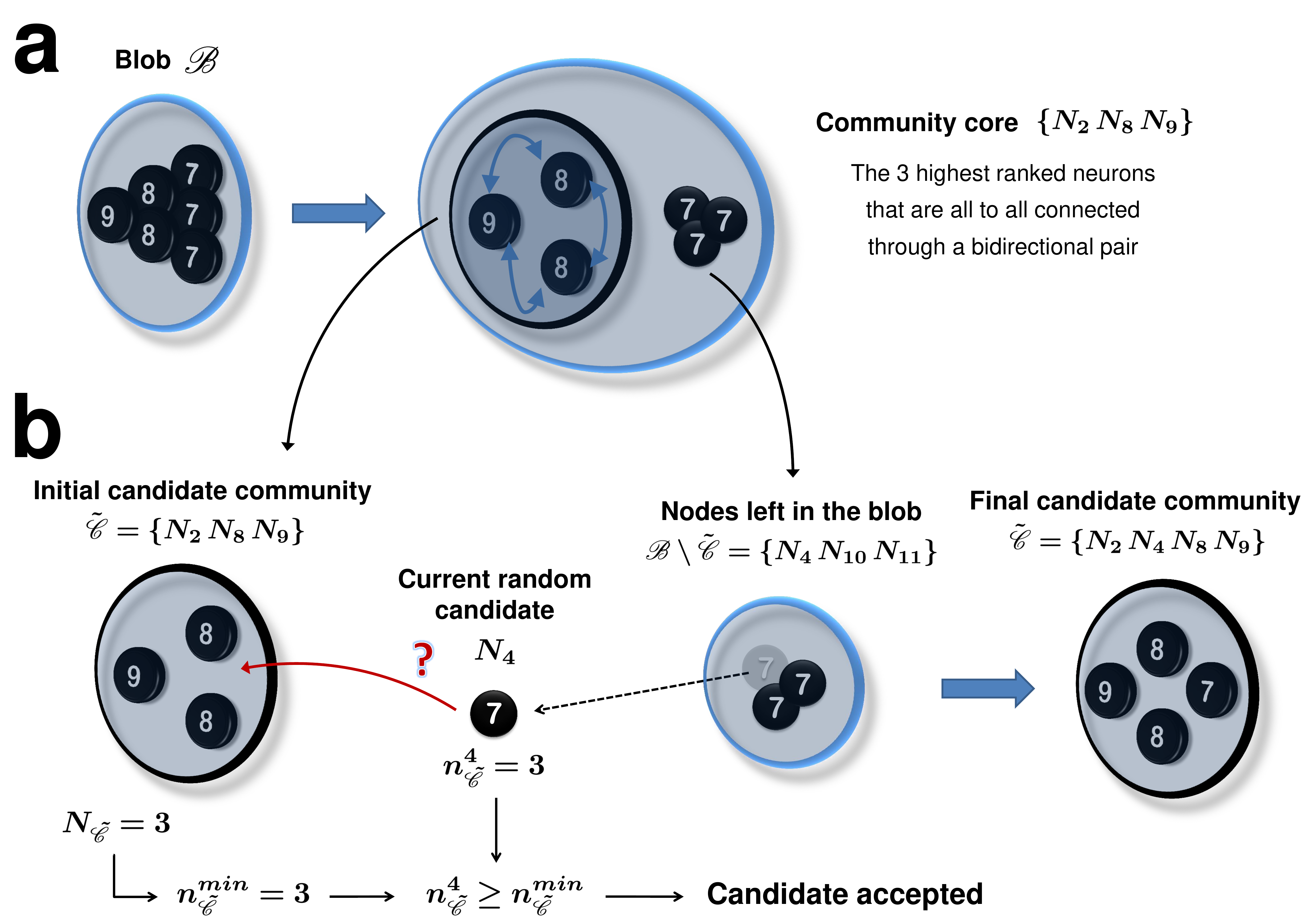}
\end{center}
\caption{{\bf Algorithm Step 3: Friendship algorithm.} \textbf{A} Detecting the candidate community core: The highest ranked neurons in the blob that are also all-to-all bidirectionally connected. \textbf{B} Building a candidate community: Starting from the core (\textit{left}), one of the neurons that are left in the blob is randomly selected at a time and, based on the community definition, its inclusion in the candidate community is evaluated (\textit{middle}). Here we show only the first iteration with the neuron $N_{4}$: Because it forms a bidirectional connection with all the $3$ neurons in $\tilde{\mathcal{C}}$ ($n^{4}_{\tilde{\mathcal{C}}} = 3$), it can be accepted, resulting in a bigger candidate community for the next iteration (\textit{right}).}
\label{fig:Algorithm_step3}
\end{figure}

In Fig. \ref{fig:Algorithm_step1}-\ref{fig:Algorithm_step3} we show the algorithm implementation on a toy network of $N = 11$ neurons, all-to-all connected and labelled as $N_{1}, N_{2}, \dots N_{11}$ (Fig. \ref{fig:Algorithm_step1}a, \textit{left}). For simplicity, instead of using $N_{i}$ when referring to the neurons of the example, we assume that indices like $i$ vary directly in the set $N_{1}, N_{2}, \dots N_{11}$. Moreover, for a better understanding, in Tab. \ref{tab:symbols} we report a list of the symbols used and their description.

\paragraph*{Step 1. Neurons popularity ranking.} From the full network's connectivity $W$ (Fig. \ref{fig:Algorithm_step1}b, \textit{left}), we derived the relative strength $Z_{ij}$ of each pair of neurons (Fig. \ref{fig:Algorithm_step1}b, \textit{middle}), given by equation (\ref{conn_pairs}), and we assess their bidirectionality using the threshold $Z_{B}$ defined in equation (\ref{bid_thresh}). This allows assigning to each neuron $i$ a number $n^{i}$ representing how many bidirectional pairs that neuron forms in the entire network (Fig. \ref{fig:Algorithm_step1}b, \textit{right}). 

Based on this information, neurons are initially entered in a bidirectional pool $\mathcal{P}$ depending on a minimum required number of bidirectional pairs $n^{min}_{\mathcal{P}}$, which can be arbitrarily chosen (Fig. \ref{fig:Algorithm_step1}a, \textit{middle}). Neurons that do not meet the \textit{pool entering condition} $n^{i} \geq n^{min}_{\mathcal{P}}$ are excluded from $\mathcal{P}$, as they do not have the basic requirement for being part of a community. As a consequence, the bidirectional pairs that these excluded neurons form in the network also cannot be part of any community, hence they should be subtracted from the $\left\lbrace n^{i} \right\rbrace$ of the involved neurons. Therefore, after $\mathcal{P}$ is formed, neurons are subject to the \textit{pool staying condition} $n^{i}_{\mathcal{P}} \geq n^{min}_{\mathcal{P}}$, where $n^{i}_{\mathcal{P}}$ is the number of bidirectional pairs formed only within ${\mathcal{P}}$. Nodes violating this inequality are excluded from ${\mathcal{P}}$ and so are their bidirectional pairs. The pool is therefore reduced and $\left\lbrace n^{i}_{\mathcal{P}} \right\rbrace$ need to be updated.  This iterative process stops when $n^{i}_{\mathcal{P}} \geq n^{min}_{\mathcal{P}}$ $\forall i \in \mathcal{P}$ or when the number of neurons left in the pool is below the noise threshold (meaning that an eventual community can be considered as a random happening, see below). In the first case, the final $\mathcal{P}$ is the working material for the next steps, whereas in the second case the entire algorithm ends with no communities found.

Differently from the following steps, nodes that are left out of $\mathcal{P}$ are definitely lost, as they will not be reconsidered again. Hence, the value assigned to $n^{min}_{\mathcal{P}}$ has to be carefully evaluated: limiting the number of neurons in the pool will greatly reduce the computational cost of the rest of the algorithm; however, the risk of not including neurons that are actually part of a community increases. Throughout this paper we adopt the "safe" choice $n^{min}_{\mathcal{P}} = 1$, for which $\mathcal{P}$ coincides with the whole network when $N$ and $Z_{B}$ are sufficiently large like the ones we use. This is also the case of the toy network we are considering in this section (all neurons of the network are admitted to the pool, see Fig. \ref{fig:Algorithm_step1}a).

Neurons in $\mathcal{P}$ can be sorted in a \textit{popularity ranking} based solely on $n^{i}_{\mathcal{P}}$ (Fig. \ref{fig:Algorithm_step1}a, \textit{right}). In doing so, nodes with the same value of $n^{i}_{\mathcal{P}}$ are treated as identical because we are (temporarily) losing all the detailed information of which pairs of neurons are effectively connected with each other. \textit{Step 2} below is built only upon the popularity ranking, therefore without the need to access this detailed information. This allows to save a considerable amount of computational resources and to speed up the research, while still obtaining great results in terms of community detection.

Popularity ranking is a preliminary step, deterministic and with no approximations (i.e. there is no loss of information) as long as the threshold $n^{min}_{\mathcal{P}}$ is kept to a low value. From now on we will be working only with the neurons in $\mathcal{P}$.

\paragraph*{Step 2. Blob search.} Our heuristic approach consists in using the popularity ranking to narrow the research to the regions in $\mathcal{P}$ where it is more likely to find a community. The key observation is that once we give a formal definition of community like equation (\ref{comm_def}), we can use it a basis for a reliable heuristic search. On the left hand side of equation (\ref{comm_def}) we have the number of bidirectional pairs formed by neuron $i$ within the community $\mathcal{C}$, which, by using the notation introduced in this section, can be rewritten as $n_{\mathcal{C}}^{i}$. As pointed out earlier, these quantities are unknown; however, they are upper bounded by $n_{\mathcal{P}}^{i}$, which corresponds to the case where all the bidirectional pairs that a neurons forms are part of the same, unique community. We can invert this argument and derive from equation (\ref{comm_def}) the size of the largest community that each neuron can potentially form within the pool:
\begin{equation}
\label{max_comm_single_neuron}
|\mathcal{C}_{i}^{max}| = \frac{n_{\mathcal{P}}^{i}}{\vartheta_{\mathcal{C}}} + 1 \quad\quad \forall i \in \mathcal{P} \,.
\end{equation}
$|\mathcal{C}_{i}^{max}|$ does not take into account which are the neurons connected with neuron $i$ through the $n_{\mathcal{P}}^{i}$ bidirectional connections. As a consequence, all the $N_{\mathcal{P}} = |\mathcal{P}|$ relations of equation (\ref{max_comm_single_neuron}) are uncoupled, differently from what happens in the community definition. Starting from equation (\ref{max_comm_single_neuron}), this second step aims at filling the gap with equation (\ref{comm_def}) by progressively incorporating these element that have been discarded, hence producing as a result sets of neurons in $\mathcal{P}$ that satisfy the full community definition. The goal of this step is indeed to find the largest possible communities in the network. This is done through a recursive two-step procedure, depicted in (Fig. \ref{fig:Algorithm_step2}): 
\begin{itemize}
\item \textit{Part 1.} Starting form equation (\ref{max_comm_single_neuron}), we find the largest set of neurons in the pool that could potentially form a community based only on the information contained in the popularity ranking. We call this a \textit{bidirectional candidate blob} $\tilde{\mathcal{B}}$. Formally, $\tilde{\mathcal{B}}$ is a set of neurons such that:
\begin{equation}
\label{blob_def}
n^{i}_{\mathcal{P}} \geq \vartheta_{\mathcal{C}} \left( |\tilde{\mathcal{B}}|-1 \right)\quad \forall i \in \tilde{\mathcal{B}} \quad\quad \text{\textit{with $\tilde{\mathcal{B}}$ as large as possible.}}
\end{equation}

Respect to equation (\ref{max_comm_single_neuron}), with equation (\ref{blob_def}) we are restoring the coupling between the equations, that is an essential feature of the community definition. On the other hand, the approximation that we are making respect to equation (\ref{comm_def}) is clear when we compare the two relations: neurons are included in $\tilde{\mathcal{B}}$ not because of the number of bidirectional connections that they form within $\tilde{\mathcal{B}}$, as the community definition would require, but depending on the total number of bidirectional connections that they form in the entire pool, $n_{\mathcal{P}}^{i}$. 

Note that $\tilde{\mathcal{B}}$ does not coincide with the potential community that the most popular neuron can form ($\max_{i \in \mathcal{P}} |\mathcal{C}_{i}^{max}|$), but it is highly likely that such a neuron is part of $\tilde{\mathcal{B}}$. In other words, the most popular neuron $i^{*} = \argmax_{i \in \mathcal{P}} |\mathcal{C}_{i}^{max}|$ is the node that has the highest probability in the entire network to belong to $\tilde{\mathcal{B}}$, hence it is the first one to be recruited. In the example, $i^{*} = N_{2}$, with $9$ bidirectional pairs formed in the pool (see Fig. \ref{fig:Algorithm_step2}a, \textit{top left}). The other neurons are organised in waves, formed by identically ranked nodes, which are evaluated one at a time in a descending order (Fig. \ref{fig:Algorithm_step2}a, \textit{middle}). At every iteration, the candidate blob is fully characterised by two quantities: the actual size $N_{\tilde{\mathcal{B}}}$ and the size $N^{max}_{\tilde{\mathcal{B}}}$ of the largest possible community the entire set $\tilde{\mathcal{B}}$ can be part of, based on the popularity ranking. Since for each neuron this is given by equation (\ref{max_comm_single_neuron}), then, for the candidate blob as a whole, $N^{max}_{\tilde{\mathcal{B}}}$ is determined by the last wave of neurons included:
\begin{equation}
\label{max_comm_blob}
N^{max}_{\tilde{\mathcal{B}}} = \min_{i \in \tilde{\mathcal{B}}} |\mathcal{C}_{i}^{max}| \,.
\end{equation}
At the beginning, $\tilde{\mathcal{B}} = \left\lbrace i^{*} \right\rbrace$, hence $N_{\tilde{\mathcal{B}}} = 1$ and $N^{max}_{\tilde{\mathcal{B}}} = |\mathcal{C}_{i^{*}}^{max}|$.  As we progressively recruit waves of neurons, $N_{\tilde{\mathcal{B}}}$ increases whereas $N^{max}_{\tilde{\mathcal{B}}}$ decreases. As long as $N_{\tilde{\mathcal{B}}} < N^{max}_{\tilde{\mathcal{B}}}$ then $\tilde{\mathcal{B}}$ can potentially be a community and we can keep on recruiting the next wave of neurons to investigate whether a larger candidate blob (which could lead to a larger community) is possible. When the inequality is no longer satisfied then the process of recruiting neurons stops. We can have two scenarios: $N_{\tilde{\mathcal{B}}} > N^{max}_{\tilde{\mathcal{B}}}$ means that some neurons in $\tilde{\mathcal{B}}$, the most recently added ones, do not have enough bidirectional pairs: even in the most optimistic case where all these pairs are within $\tilde{\mathcal{B}}$, these neurons do not meet the community threshold, given the actual size of $\tilde{\mathcal{B}}$. Thus, the largest possible candidate blob is the one found at the previous iteration. This is the case of the toy network we are considering, as shown in Fig. \ref{fig:Algorithm_step2}a, \textit{top right}. The second scenario is when $N_{\tilde{\mathcal{B}}} = N^{max}_{\tilde{\mathcal{B}}}$, and the largest possible candidate blob is the set of neurons found at the current iteration.

\item \textit{Part 2.} At this stage we have a set of neurons, a candidate blob $\tilde{\mathcal{B}}$ (Fig. \ref{fig:Algorithm_step2}b, \textit{left}), that satisfies equation (\ref{blob_def}): according to the number of bidirectional connections that each of them forms in the pool, $\tilde{\mathcal{B}}$ is suitable to form a community. We can therefore move on and restore the full community definition, equation (\ref{comm_def}), by computing the number of bidirectional connections $n^{i}_{\tilde{\mathcal{B}}}$ that each neuron in $\tilde{\mathcal{B}}$ forms within $\tilde{\mathcal{B}}$ itself (Fig. \ref{fig:Algorithm_step2}b, \textit{middle}). In other words, we are substituting $n_{\mathcal{P}}^{i}$ with $n^{i}_{\tilde{\mathcal{B}}}$ in equation (\ref{blob_def}), obtaining exactly the community definition. Thus, if each neuron in $\tilde{\mathcal{B}}$ satisfies the condition, then there is a community structure and we call it a \textit{bidirectional blob} $\mathcal{B}$ (Fig. \ref{fig:Algorithm_step2}b, \textit{right}): a set of neurons that certainly contains at least one bidirectional community. 

If the community definition is not satisfied, then we withdraw those neurons that violate it to obtain a new $\tilde{\mathcal{B}}$ and to apply \textit{Part 2} again. This refinement process continues until the algorithm finds a blob $\mathcal{B}$ or until $\tilde{\mathcal{B}}$ contains only one neuron, meaning that there is no blob. 
\end{itemize}

Whenever this step gives a non empty blob as a result, then we proceed with \textit{Step 3} below to finally find the communities that are present in $\mathcal{B}$. After this, we temporarily eliminate from the pool all the neurons that have been detected as being part of a community so far, and to this modified pool we apply \textit{Step 2} from the beginning.  Therefore, if a neuron is found to be part of a community, it does not get the chance to be evaluated again for being included in other blobs, meaning that blobs are all disjointed sets. This is one of the reasons why we introduce \textit{Step 3} below, which is built to detect overlapping communities.

The procedure continues until there is no blob found. In this case the entire algorithm goes to an end and its final outcome are all the bidirectional communities found so far within the previously detected blobs. Thus, the result of this step is a set of non-overlapping blobs, each of them containing for sure at least one bidirectional community.

\paragraph*{Step 3. Looking into a blob: the friendship algorithm.}
The set $\mathcal{B}$ that we obtain at the end of \textit{Step 2} certainly forms a community, as it satisfies the definition. However: \textit{i)} two different communities, or at least parts of them, may be detected within the same blob (\textit{resolution problem}). \textit{ii)} The method does not correctly identify overlapping communities. If there is an overlap between any of them, the above procedure assigns the intersection region to only one community (\textit{overlapping problem}). \textit{iii)} Mistakes may occur during the identification process: some neurons that are originally part of a community may have been left out of the blob that contains that community, we call them \textit{good friends}, whereas some other neurons may have been erroneously included in $\mathcal{B}$, we call them \textit{false friends} (\textit{accuracy problem}).

In order to address these issues, at this stage we define a procedure that builds up a community neuron by neuron through direct verification of the definition. This approach is much more feasible and robust at this stage rather than at the very beginning, for two reasons: \textit{i)} we apply it to a very limited set of neurons, i.e. the blob $\mathcal{B}$, and \textit{ii)} we already know that these neurons are relevant in terms of community structure.

\textit{Step 3} starts by selecting the three neurons of the blob that are the most popular ones and at the same time form only bidirectional pairs among them. We call this \textit{candidate community core}, and its purpose is to give a likely basis where to start building the \textit{candidate community} ${\tilde{\mathcal{C}}}$ (Fig. \ref{fig:Algorithm_step3}a). The reason why we choose three neurons is to minimise mistakes: there is no guaranty that these three nodes really belong to the same community, but the probability of this event to happen is high, and it monotonically decreases with the size of the candidate community core. Also, among all the possible triplets of neurons, these three nodes maximise the probability of having three neurons being part of the same community, since they are the highest in the ranking. On the other side, the choice of having a core of only two neurons does not seem very reliable because, due to the random generation process of connections in the network (see below), the event that two neurons are connected with a bidirectional pair is very common, even among neurons that are not part of any community. Once we identify the candidate community core, in each iteration we randomly select a neuron in the blob and we ask if it satisfies the community definition (Fig. \ref{fig:Algorithm_step3}b): given the current size of the candidate community $N_{\tilde{\mathcal{C}}}$, by using the definition equation (\ref{comm_def}) and rounding up the result, we have the minimum number of bidirectional connections $n^{min}_{\tilde{\mathcal{C}}}$ that the \textit{candidate neuron} needs to form with the current members of ${\tilde{\mathcal{C}}}$ in order to join it.

In Fig. \ref{fig:Algorithm_step3}b we show the procedure for the first iteration only, where $\tilde{\mathcal{C}} = \left\lbrace N_{2} \, N_{8} \, N_{9} \right\rbrace$, hence $N_{\tilde{\mathcal{C}}} = 3$, $n^{min}_{\tilde{\mathcal{C}}} = 3$, and $N_{4}$ is the randomly selected node. This neuron forms $7$ bidirectional connections in the entire network, but at this stage this is not relevant anymore. What matters is that it forms a bidirectional pair with each of the neurons in the current candidate community ($n^{4}_{\tilde{\mathcal{C}}} = 3$), meaning that it is "friend" with all of them and thus it can clearly be accepted in $\tilde{\mathcal{C}}$. In the next iteration, the candidate community is then $\tilde{\mathcal{C}} = \left\lbrace N_{2} \, N_{4} \, N_{8} \, N_{9} \right\rbrace$, which results in $N_{\tilde{\mathcal{C}}} = 4$ and $n^{min}_{\tilde{\mathcal{C}}} = 3$ again. Thus, the neuron that will be selected, either $N_{10}$ or $N_{11}$, needs to form at least $3$ bidirectional connections with the $4$ candidate community's members in order to join it. This is exactly what happens in this example, and, since in the last iteration also the last neuron turns out to have enough friends in $\tilde{\mathcal{C}}$ (for which it will be $N_{\tilde{\mathcal{C}}} = 5$ and $n^{min}_{\tilde{\mathcal{C}}} = 4$), the final candidate community will coincide with the blob: $\tilde{\mathcal{C}} = \left\lbrace N_{2} \, N_{4} \, N_{8} \, N_{9} \, N_{10} \, N_{11} \right\rbrace$. 

Note that in early iterations, when the candidate community is not well formed yet, false friends still get a chance to pass the test and be recruited in $\tilde{\mathcal{C}}$. On the other hand, strongly connected neurons within the original community have much higher chances to pass the test, no matter at which iteration they are selected. Thus, once the recruitment within the blob has finished and $\tilde{\mathcal{C}}$ is formed, we check again that each neuron is entitled to stay in $\tilde{\mathcal{C}}$ through the verification of the community definition (\textit{false friends expulsion}). In the example we are using, no neuron in $\tilde{\mathcal{C}}$ fails to meet the requirement so they all definitely earn the right of being in the candidate community.

Most likely, the above procedure returns a set that is a single, well consistent candidate community, hence solving the resolution problem. However, it might not be the original complete community: good friends may be left out, most likely in the blob $\mathcal{B}$ but also in the pool $\mathcal{P}$. Therefore, we start recruiting for a second time with the condition of satisfying the community definition, first among the neurons left in the blob and then within the pool (\textit{good friends inclusion}). This procedure, together with the false friends expulsion, addresses the accuracy problem. Moreover, recruitment among neurons in the pool allows to consider for the current candidate community also neurons that have been previously included in other blobs or communities, hence solving the overlapping problem. In our example, there are no neurons left in the blob so we check the ones in the pool: $N_{1}$, $N_{3}$ and $N_{7}$ are recruited whereas $N_{5}$ and $N_{6}$ are left out, giving finally a community candidate of $9$ neurons that coincides with the original true community.

\paragraph*{Step 4. Candidate community global control.}
As pointed out earlier, based on Esposito et al. (2014)\cite{Esposito2014}, for a set of neurons to be a bidirectional community $\mathcal{C}$, its symmetry measure equation (\ref{sym_meas}) must exceed a threshold value $s_{B}$, which depends on the distribution of connections considered. Now that we isolated a candidate community from the rest of the network, we are in the position of applying this criterion. Candidate communities that do not pass this test are sets that cannot be qualified as communities. Note, however, that they are still bidirectional communities in the sense of our topological definition equation (\ref{comm_def}). Since this definition is threshold-based, it introduces a binary criterion with subsequent loss of information. The definition is guidance for community detection that reduces the weighted network to an un-weighted one. Thus, once the research has been successful, the complete information stored in the weights needs to be recovered and the actual identity as bidirectional community can be finally evaluated by means of the symmetry measure. 

Sets that cannot be qualified as communities present an excess of bidirectional pairs due to the random generating process, which made these sets to be detected as possible communities, but failure in the symmetry measure test means that the rest of the pairs are far from being bidirectional, hence pulling the value of the symmetry measure down within the randomness boundaries. This is no evidence that learning took place in the specific set as a whole. Statistically, this situation is likely to happen for small sets of neurons, and indeed this is when we observe failure of the symmetry measure test. These sets of neurons are therefore safely withdrawn. 

\paragraph*{Step 5. Noisy candidate communities identification.}
Besides the real communities that are present in the network as a result of learning, communities can be also formed out of chance, due to the randomness in the network's connectivity: it is highly likely that small sets of $4-5$ neurons show community properties and thus will be detected as such. Since the probability of randomly forming a community dramatically drops with the size, we can define a \textit{noise threshold} $\vartheta_{noise}$ and discard all sets below such a threshold. This clearly fixes a lower limit to the resolution of the algorithm. However, the maximum size of a random community, which ideally corresponds to such a threshold, grows with the size of the network in a way that is sub-linear, allowing to set a unique, relatively small threshold for all the networks with a large size that does not affect the overall performance.

\paragraph*{Step 6. Single community reduction.}
At this stage we have single communities $\mathcal{C}$, but we might have a final \textit{redundancy problem}, especially for networks of small size: it can happen that the same original community has been detected more than once, every time with different false friends included and good friends left out. It can also happen that the original community is broken into two overlapping parts detected as different communities. The final step is therefore trying to resolve these issues calculating the overlap degrees between each pair of communities. Pairs with an overlap exceeding the \textit{overlap threshold} $\vartheta_{\omega}$ are merged together and the symmetry measure is used as an evaluating criterion: if the value on the merged community is higher than the values of the single communities, then the merged community definitely replaces the two single ones, otherwise they are kept separate.

\subsection*{Network and communities generation: benchmark procedure}
To test the above algorithm, we generated \textit{in silico} data representing several different scenarios, which will be discussed in the Results section alongside the algorithm performance. Below, we describe the general procedure used to produce a connectivity matrix for a network containing bidirectional communities.

The first step for creating a bidirectional community is to decide the value of its symmetry measure, which has to be in the range $\left[s_{B}, 1\right]$ \cite{Esposito2014}. By definition, this gives the mean value of the strength of the connection pairs in the community, $\left\langle Z \right\rangle = 1 - s$. Because the learning process shapes the connections of a community in the same direction, it is reasonable to assume that at the end of learning the pairs form a Gaussian distribution. Therefore, for each community in the network, we generated the set of $\left\lbrace Z_{ij}\right\rbrace$ according to a Gaussian distribution with mean $\left\langle Z \right\rangle$ and standard deviation $\sigma$, which is a free parameter. We recall that $Z$ is a variable ranging from $0$ to $1$ and that the bidirectionality region is $\left[0, Z_{B}\right]$. Based on this, two issues may arise when we generate the pairs, related to the two boundaries and to the choice of $\sigma$: \textit{i)} some of the $Z_{ij}$ could be negative. If this is the case, the tails of the distribution are symmetrically folded towards the inside so as to guarantee the non-negativity of the $\left\lbrace Z_{ij}\right\rbrace$ and to preserve the mean value of the distribution. \textit{ii)} A considerable part of the distribution could fall in the randomness domain ($Z > Z_{B}$), meaning that many pairs will not be classified as bidirectional. As a consequence, some neurons may not form the minimum number of bidirectional pairs required by the community definition, resulting in the whole set not being a community anymore. To avoid this issue, we make sure that the integral of the Gaussian in the bidirectional region, which gives the probability of forming a bidirectional pair $\wp_{B}$, exceeds the community threshold $\vartheta_{\mathcal{C}}$.

Once we have the set of $\left\lbrace Z_{ij}\right\rbrace$ for each community, we can generate the single connections $w_{ij}$. A first half of them is directly drawn from the uniform distribution in $\left[ 0, 1\right]$, be the upper (or the lower) triangular part of the community's connectivity matrix. This first half, together with the $\left\lbrace Z_{ij}\right\rbrace$, is used to compute the second half of the single connections by means of equation (\ref{conn_pairs}). The rest of the connections in the network are drawn from the uniform distribution in $\left[ 0, 1\right]$. 

Overlaps between communities are governed by the set of parameters $\left\lbrace \omega_{\lambda\rho} \right\rbrace$ representing the fraction of the community $\rho$ that is in common with the community $\lambda$:
\begin{equation}
\omega_{\lambda\rho} = \frac{|\mathcal{C}_{\lambda} \cap \mathcal{C}_{\rho}|}{|\mathcal{C}_{\rho}|}
\end{equation}
We allow overlaps only between subsequent pairs of communities. In other words, we can progressively enumerate the communities in the network in such a way each of them overlaps at most with only the previous and following one. Formally: $\omega_{\lambda\rho} = 0$ if $|\lambda - \rho| \geq 2$, leading to a tridiagonal matrix of overlaps. In cases of overlap between two communities, after having generated the first community, the mean of the pairs in the intersection is computed and it is used to offset the mean of the Gaussian distribution for the rest of the pairs in the second community, so as to preserve the value of the symmetry measure that we chose.

The set of parameters $\left\lbrace s_{\lambda} \right\rbrace$, $\left\lbrace \sigma_{\lambda} \right\rbrace$, $\left\lbrace \omega_{\lambda\rho} \right\rbrace$ we introduced here for the connections generation, together with the size of communities $\left\lbrace N_{\lambda} \right\rbrace$ and network $N$, entirely define the structure of a network, but they do not uniquely determine its connectivity because all the connections are generated through the above mentioned random process. Due to the presence of random elements in both data generation and detection procedure, for each combination of parameters we consider, we repeat the experiment $n_{iter} = 100$ times. Each experiment, or run, consists in generating the network connectivity as described above and applying our detection algorithm. Cumulative and averaged results are displayed in the appropriate section.

\subsection*{Analysis of the results: measuring successful detection}
At the end of each run, on one side we have the communities that we generated at the beginning, i.e. the real communities, and on the other side those detected by the algorithm. One way of measuring the quality of the results is to count how many good neurons have been detected. However, since this is going to be displayed as an average over the runs, we may lose too much information about the single run. Also, as pointed out earlier, failure in detecting single neurons may still happen despite the bulk of the community has been correctly identified.

Therefore, we introduce a criterion to determine successful community detection: whenever the number of neurons in a detected community equals at least a fraction $\vartheta_{recog}$ of the neurons in a real community, we count that real community as successfully identified. If there is more than one detected community for which this happens relatively to the same real community, then the one with the highest percentage is considered to be one matching the real community and the others are counted as false communities, unless they result to match some other real community in the network. We choose $\vartheta_{recog} = 75\%$ as in our opinion three quarters is a fraction that already carries the distinctive features of the community to which it belongs.

At the end of the results' evaluation, the analysis of the algorithm performance can be done by using he following information on each real community: \textit{i)} how many times it has been successfully detected in all runs, and \textit{ii)} how many good neurons have been identified as average across the runs. Alongside, we also display information about false communities that have been detected and false neurons included in good communities. Results about communities' detection provide a quantitative tool to evaluate the goodness of the algorithm, whereas neurons detection provide a qualitatively information on its accuracy. Finally, we show the time needed to run the algorithm.

\subsection*{Thresholds} The algorithm described above makes use of $5$ customisable thresholds, see Tab. \ref{tab:symbols}. Throughout this paper we keep them fixed at their respective values. Since we assume a uniform distribution of connections prior learning, we can directly rely on the results of Esposito et al. 2014 \cite{Esposito2014} for uniform distributions: by fixing a level of confidence at $p=0.05$, the bidirectionality threshold we use is $s_{B} = 0.6954$, which in turn gives $Z_{B} = 0.3046$. The threshold $\vartheta_{\mathcal{C}}$ for community existence is rather arbitrary and it can be fixed according to how dense we require the communities to be. In the present study we choose $\vartheta_{\mathcal{C}} = 75 \%$. Concerning the noise effect, after observing the size of the noisy communities detected by the algorithm, we fix $\vartheta_{noise} = 30$.  For the other thresholds, also arbitrary, we use $n^{min}_{\mathcal{P}} = 1$ as a safe choice (as previously stated) and $\vartheta_{\omega} = 25\%$ as a limit case before two communities can be considered as part of a single bigger one (after evaluation of symmetry measure, see \textit{Step 6}).

\section*{Results}

In this section we present the results obtained by applying the community detection algorithm to networks of neurons with different community structures. In all the cases, we assume that the given network is the final product of a learning process that shaped the connections of some sub-regions away from the initial uniform distribution, to form what we called bidirectional communities, equation (\ref{comm_def}). The rest of the connections remain unchanged and therefore they are uniformly distributed. Network connectivity is generated according to the procedure outlined in Methods section.

Since the learning process is not explicitly simulated here, we have total control on the final structure of the network, through the tuning of $6$ sets of parameters: the size of the network $N$, the number of communities $\nu$, the size of communities $\left\lbrace N_{\lambda} \right\rbrace$ and the overlap between communities  $\left\lbrace \omega_{\lambda\rho} \right\rbrace$ define the architecture of a network. The symmetry measure of the communities  $\left\lbrace s_{\lambda} \right\rbrace$ and the standard deviation of the connection pairs in the communities  $\left\lbrace \sigma_{\lambda} \right\rbrace$ define how much a community has been shaped towards bidirectionality.

The way the algorithm is constructed, we expect that strong bidirectional communities, i.e. with $s \to 1$ and small $\sigma$, are the easiest to detect, compared with bidirectional communities with $s \to s_{B}$ and large standard deviation. The degree of difficulty in the detection of a community can be derived from the way we generate the community itself. Indeed, as pointed out in the Methods section, $s$ and $\sigma$ determine the number of bidirectional pairs that each neuron forms in the community, through a random process. Thus, necessary condition for a set of neurons to be a community (and therefore to be detected) is that this number exceeds the threshold for community existence. Also, the closer this number to the threshold (from above), the more difficult the detection of the community.

Each simulation consists of $n_{iter} = 100$ runs, each of them starts with communities and network generation, continues with the communities' detection algorithm and finally ends with the evaluation of the results, where we compare detected and generated communities. To evaluate the algorithm performance, we use two indicators for each community generated in the network. The first quantity counts how many times a community has been successfully detected during the $n_{iter}$ iterations (see Methods). Once a community has been correctly identified, the second indicator measures how many nodes of the generated community have been detected, and displays this information as an average percentage across $n_{iter}$ iterations and relative to the total number of nodes in the generated community. Additional indicators for the number of false communities detected and for the percentage of false neurons in a correctly detected community complete the evaluation.

\subsection*{Networks with a single community}
Alongside the size of the community, we introduce the community to network ratio $r_{c/n} = N_{\mathcal{C}} / N$, which is a more significant indicator to assess the algorithm performance. A complete evaluation (at least in the single community case) requires, therefore, carrying out $3$ different analyses, corresponding to fixing one of the three quantities $N_{\mathcal{C}}, N, r_{c/n}$ while varying the other two.

\begin{figure}[h!t]
\begin{center}
\includegraphics[width=1\linewidth]{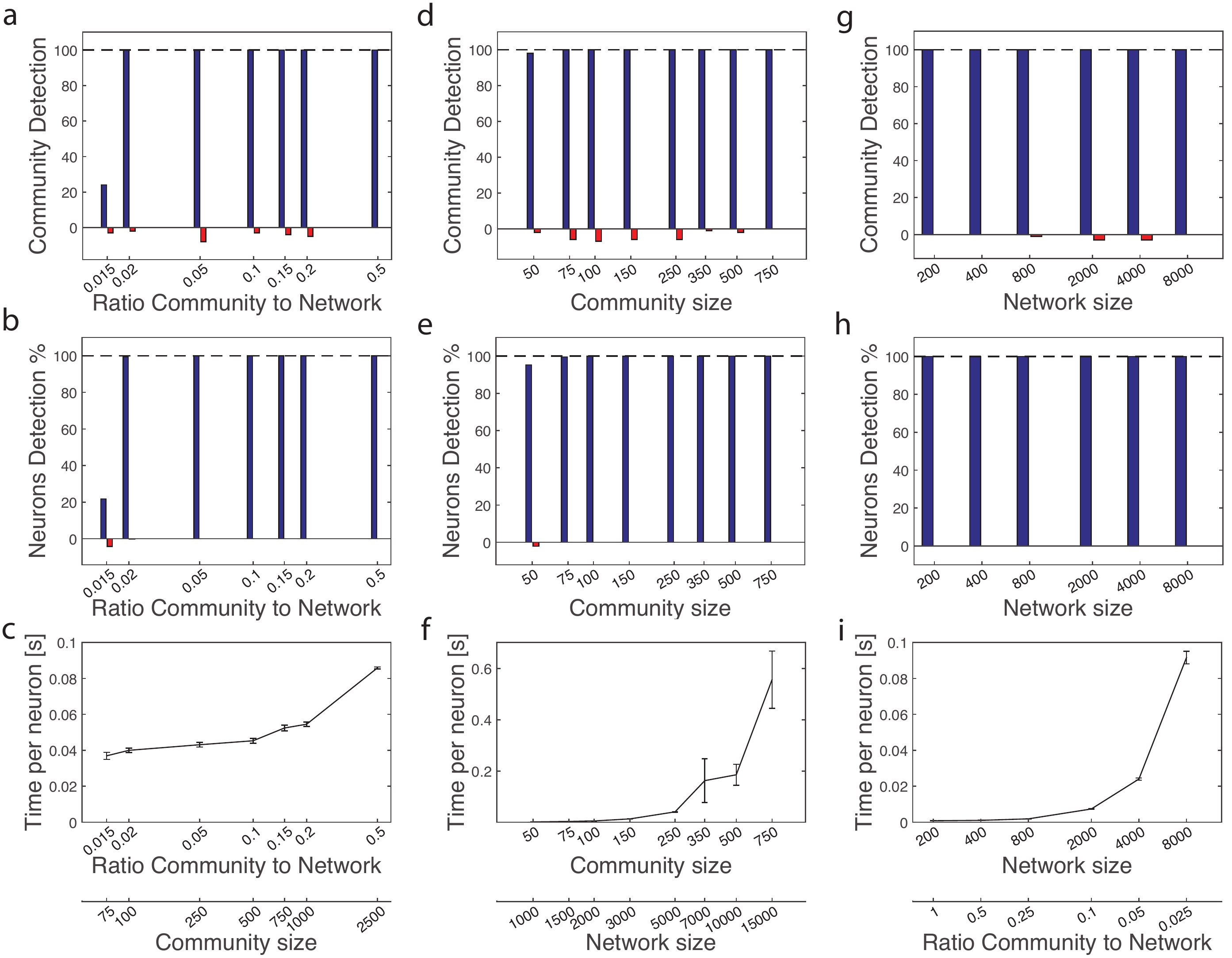}
\end{center}
\caption{{\bf Algorithm Performance for the case of single community of $\bm{s_{\mathcal{C}} = 0.75}$ and $\bm{\sigma_{\mathcal{C}} = 0.05}$ embedded in a network.} \textbf{A - C} Network' size fixed at $N = 5000$ neurons while varying community' size. \textbf{D - F} Ratio community to network fixed at $r_{c/n} = 0.05$ while varying both network' and community' size. \textbf{G - I} Community' size fixed at $N_{\mathcal{C}} = 200$ neurons while varying network' size. All results in each panel are relative to $n_{iter} = 100$ repetitions. \textbf{A, D, G} Cumulative community detection. \textit{Blue bars:} Successful detection, \textit{Red upside down bars:} False detection. The dashed line represents the best possible performance of correctly detecting the community all the time. \textbf{B, E, H} Average percentage of neurons detection, relative to the size of the generated community. \textit{Blue bars:} Good neurons, \textit{Red upside down bars:} False neurons. \textbf{C, F, I} Simulation time per neuron in the network. Error bars represent standard error. Note that the scale of all x-axes is logarithmic.}
\label{fig:Results_Single_community}
\end{figure}

\subsubsection*{Single, size-varying community in a network of fixed size}
We begin the analysis of the algorithm performance with the most common and typical problem: given a network, we want to know if there is a community and of which size. Therefore, we fix the size of the network at $N = 5000$ neurons and we systematically vary the size of the community in the set $\left\lbrace 75,100,250,500,750,1000,2500 \right\rbrace$. The community is generated with $s_{\mathcal{C}} = 0.75$ and $\sigma_{\mathcal{C}} = 0.05$, resulting in a probability of forming a bidirectional pair $\wp_{B} = 0.863$. Correctly, $\wp_{B} > \vartheta_{\mathcal{C}}$ (see Network and community generation subsection in Methods).

In Fig. \ref{fig:Results_Single_community}a we report the results concerning community detection: blue bars show the cumulative number of successful detection of the generated community, whereas the upside down red bars count the number of false communities. Similarly,  Fig. \ref{fig:Results_Single_community}b shows the average percentage of good neurons (blue bars) and false neurons (upside down red bars) in the detected community. Blue bars carry what we can call a \textit{positive information} as we want to maximise them, whereas red bars is what we want to minimise to zero, hence they carry a \textit{negative information}. The horizontal black dashed line marks the optimality level for positive information: when for the same value of $r_{c/n}$ both bars of Fig. \ref{fig:Results_Single_community}a,b hit this level means that the algorithm has detected all the neurons forming that community all the time. For the values considered here, this is almost always the case, except for the smallest community case where detection of the community is successful only $20\%$ of the times. Note that whenever this community is identified, the algorithm correctly recruits all the good neurons (the blue leftmost bars of \ref{fig:Results_Single_community}a,b have the same height). As expected, these results suggest that the bigger the size of the community the easier to detect it, with a critical value of $\sim 100$ neurons.

Fig. \ref{fig:Results_Single_community}c shows a very interesting result, direct consequence of the algorithm architecture: the computational time per neuron in the network is smaller for small communities. In other words, when we increase the difficulty of the task, the time needed for the detection is reduced, provided that the size is above the critical value for the search to be successful. It is also interesting to note that if we increase the bidirectionality of the community (by increasing the value of its symmetry measure), the algorithm time is the same (result not shown here). This means that detection time is not affected by the internal structure of the community but only by its size.

\subsubsection*{Single, size-varying community in a size-varying network with fixed ratio}
The above scenario gives only partial information on the goodness of the algorithm, as the size of the network is fixed to a single value: Fig. \ref{fig:Results_Single_community}a-c show results when we vary only the size of the community to account for different ratios community to network. However, the task of finding a community of $100$ neurons in a network of $1000$ nodes might be different from searching for a community of $1000$ in a network of $10000$ nodes.

Therefore, we investigate a second scenario: a single, size-varying community within a network whose size also varies, in such a way to keep the ratio community to network fixed. We choose a relatively small value $r_{c/n} = 0.05$, while the size of the community varies between the values $\lbrace 50,75,100,150,250,350,500,750 \rbrace$. The size of the network varies accordingly from $1000$ to $15000$. As above, $s_{\mathcal{C}} = 0.75$ and $\sigma_{\mathcal{C}} = 0.05$.

Results are shown in Fig. \ref{fig:Results_Single_community}d-f, with the same meaning of quantities and colours as in Fig. \ref{fig:Results_Single_community}a-c. Detection is perfect almost all the time, with very few mistakes mostly in the sense of detecting false communities. At first, the performance is fairly independent of the absolute sizes, as expected. A more careful inspection shows that slightly better performances are obtained for larger sizes. The reason could be that for larger networks the fluctuations on the bidirectional pairs formed out of chance become smaller and for the neurons being part of the generated community is easier to stand out of the crowd of nodes, hence the precision of the algorithm increases.

\subsubsection*{Single, fixed size community in a size-varying network}
Finally, to complete the analysis of the single community case, we study the algorithm performance when we increase the size of the network while keeping fixed the number of nodes in the community. We choose a small community of $N_{\mathcal{C}} = 200$ neurons and we vary the size of the network in the set $\lbrace 200,400,800,2000,4000,8000 \rbrace$. The ratio community to network varies accordingly from $1$ to $0.025$. Again, $s_{\mathcal{C}} = 0.75$ and $\sigma_{\mathcal{C}} = 0.05$. 

Results are shown in \ref{fig:Results_Single_community}g-i. Once again, the performance of the algorithm is excellent in the range of values considered, in terms of both positive and negative information. In particular, \ref{fig:Results_Single_community}h shows that the algorithm finds exactly the $200$ neurons forming the community all the time, with no false neurons. As expected, increasing the size of the network also increases the time needed for the detection, with dependence from the time per neuron of the network that looks quadratic.

\begin{table}[h!t]
\begin{center}
\begin{tabular}{c|ccccc}
\toprule
Community	&$N_{\mathcal{C}}$	&$s_{\mathcal{C}}$	&$\sigma_{\mathcal{C}}$	&$\omega_{\mathcal{C}\mathcal{C}'}$	&$\wp_{B}$\\
\midrule
1			&200		&0.75		&0.05			&-				&0.86\\
2			&200		&0.75		&0.05			&0.2			&0.86\\
3			&500		&0.74		&0.05			&0.1			&0.81\\
4			&150		&0.74		&0.05			&0.2			&0.81\\
5			&150		&0.79		&0.1			&0				&0.81\\
\bottomrule
\end{tabular}
\end{center}
\begin{small}
\textbf{\refstepcounter{table}\label{tab:multicomm_param}Table 4.\arabic{table}}{ List of parameter's values used to generate the community structure in the case of $\nu = 5$ communities. \textit{Column 1:} Community progressive number. \textit{Column 2:} Size of the community. \textit{Column 3:} Symmetry measure. \textit{Column 4:} Standard deviation of the connection pairs $Z$. \textit{Column 5:} Overlap with the previous community, expressed as number of common neurons divided by the number of total neurons in the community. \textit{Column 6:} Probability that a neuron of the community forms a bidirectional pair, as a result of a Gaussian distribution with parameters based on Columns 3 and 4.}
\end{small}
\end{table}

\begin{figure}[h!t]
\begin{center}
\includegraphics[width=1\linewidth]{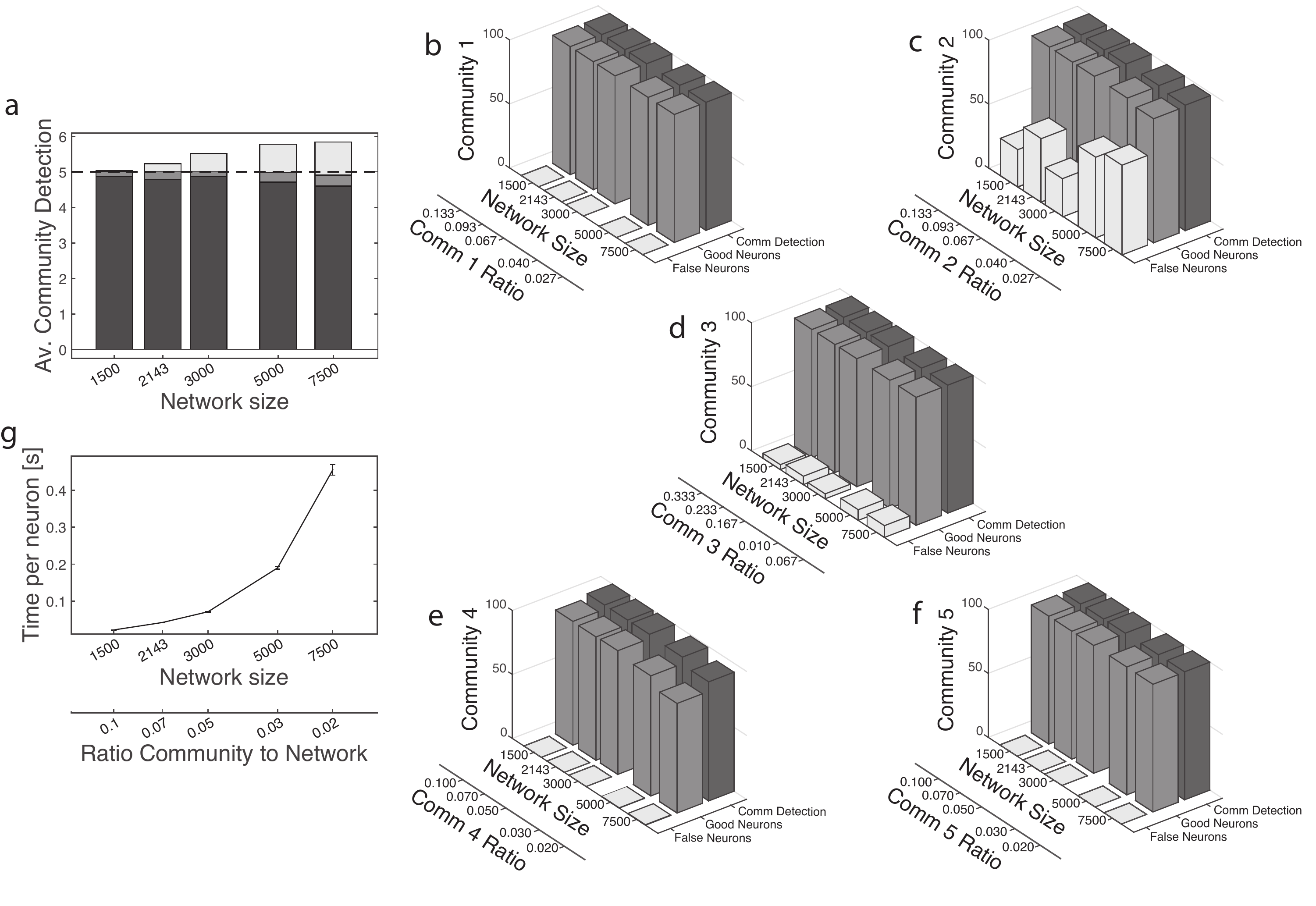}
\end{center}
\caption{{\bf Algorithm Performance for a network with complex structure.} Five communities with different sets of parameters, see Tab.~\ref{tab:multicomm_param}, are embedded in the network. Communities' size are kept fixed while varying network' size. All results in each panel are relative to $n_{iter} = 100$ repetitions. \textbf{A} Global performance. \textit{Dark grey bars:} Successful detection with all communities resolved, \textit{grey bars:} Successful detection with two communities unresolved (see text for details), \textit{light grey bars:} False communities. The dashed line represents the best possible performance of correctly detecting all the five communities all the time. \textbf{B - F} Single community detection statistics. \textit{Dark grey bars:} Cumulative community detection over the $100$ repetitions, \textit{grey bars:} Good neurons, \textit{light grey bars:} False neurons. Note that the (expected) discrete amount of false neurons detected in communities $2$ and $3$ is due to the unresolved cases between these two communities (see text for a discussion). \textbf{G} Simulation time per neuron in the network. Error bars represent standard error. Note that the scale of all x-axes is logarithmic. The ratio community to network below panel g is relative to the smallest community in the network.}
\label{fig:Results_5_communities}
\end{figure}

\subsection*{A multiple communities case}
Finally, we wish to study the behaviour of our detection algorithm when more than one community is present in the same network. As an example, we choose a challenging task: a network with $5$ communities generated with different parameters' values so as to have a certain complexity in the overall structure, see Tab. \ref{tab:multicomm_param}. The size of network is varied in the set $\lbrace 1500,2143,3000,5000,7500 \rbrace$. Note that values of the symmetry measure are all very close to the limit between bidirectionality and randomness, making the detection more difficult, as can be inferred from the last column of the table. 

Fig. \ref{fig:Results_5_communities}a shows the global performance of the algorithm, in the form of stacked bars for each value of the network's size considered. The \textit{dark grey} part at the bottom of the bars counts how many times the $5$ communities have been correctly detected as $5$ different communities, as an average over $n_{iter} = 100$ runs. The central part in \textit{grey} shows the average number of times that a community has been detected as an unresolved community, i.e. two overlapping communities detected as a single big one (never more than two). The upper part in \textit{light grey} shows the average number of false communities. Clearly, as the size of network increases, the number of false communities also increases, but the performance on the $5$ true communities remains stable and optimal. Indeed, all the communities are detected almost all the time, either resolved (more than $95\%$ of the time) or unresolved. From a more detailed analysis of the results, it can be seen that when there is an unresolved community this is always the union of the communities $2$ and $3$ in Tab. \ref{tab:multicomm_param}. This is reasonable as $\mathcal{C}_{3}$ is by far the largest in the network and the overlap of $0.1$ with community $\mathcal{C}_{2}$ means that they have $50$ neurons in common. From the perspective of community $2$ this is a considerable overlap of $25\%$, which indeed equals the value we chose for the overlap threshold $\vartheta_{\omega}$. It is therefore a matter of few neurons whether these two communities are merged or not during the last step of the algorithm (see Methods).

Fig \ref{fig:Results_5_communities}b-f shows the algorithm performance community by community, with bars showing the cumulative community detection (\textit{dark grey}), percentage of good neurons (\textit{grey}) and percentage of false neurons (\textit{light grey}). Detection of communities $1$ and $5$ is perfect, both in terms of good neurons and false neurons. Communities $2$ and $3$ present a visible amount of false neurons, most of which, however, are due to the unresolved cases between these two communities. In this sense they are not completely false neurons. Finally, community $4$ is the only one showing a decrease in the performance as the size of the network increases, with percentages still above $85\%$ for $N = 7500$. The reason is that $\mathcal{C}_{4}$ is the most difficult to detect because is the one with the lowest values of $N_{\mathcal{C}}$, $s_{\mathcal{C}}$ and $\sigma_{\mathcal{C}}$.

The last panel, Fig. \ref{fig:Results_5_communities}g shows the simulation time per neuron in the network. Note that, compared to the case of a single community, the time needed is nearly five times larger, suggesting that it grows linearly with the number of communities.

\section*{Discussion}

In this paper we address the problem of structure detection in networks of neurons, which is of crucial importance in the study of connectome in Neuroscience, framing it within the well-known, in Network Theory, community detection problems. By nature, networks of neurons are weighted and directed graphs, which makes the problem of structures searching in these networks one of the most difficult ones to approach. In the general, for most of the problems related to structures detection in Graph and Networks Theory it is not possible to give an exact solution, hence heuristic algorithms are often adopted.

Here we present an algorithm for the detection of a particular class of communities in large scale simulated networks. Thus, this is intended mainly as a tool to help \textit{in silico} research aiming at understanding the connectome. In the future, the opportunity of having direct access to synaptic weights, and therefore to connectivity matrices, may also allow a direct application of the algorithm to experimental data. Moreover, the algorithm could be of more general interest for pure studies in Networks and Graph Theory and it can be adapted to similar problems in other disciplines where nodes are not neurons.

Differently from cliques, that are very well defined objects, the concept of community is vague and we cannot find a unique definition in the literature. Traditional methods for community searching are either based on cliques or define a community \textit{a posteriori}, i.e. as the outcome given by the algorithm. Here we propose a different approach by giving a formal definition for communities \textit{a priori}. We show how having such a definition is an advantage as the algorithm can use it directly for a more efficient and direct research. Also, the method is based on a definition of symmetry measure, which allows manipulating the original connectivity and deal with quantities carrying much reduced information. This implies a loss of information, but we show that results are excellent and there is a great benefit in terms of time and computational resources.

The algorithm we present is based on statistics and it requires that the distribution of the connections is known and is somehow regular: we assumed a uniform distribution, but in principle the method works for any kind of distribution for which it makes sense to define mean and variance. As such, differently from traditional approaches, our method works better for large number of neurons: we show that increasing both the size of the network and the community gives better results. Also, we chose to focus on bidirectional communities, but the procedure can be extended to other kind of communities, for instance unidirectional ones. Indeed, similarly to bidirectional structures, experimental results show also an excess of unidirectional connections in some parts of the brain \cite{Wang2006,Lefort2009,Pignatelli2009}. Generalisation to sparse networks should be also possible.

The results we present here are relative to worst case scenarios, because the communities are generated with symmetry measure very close to the random domain. For communities that are more markedly bidirectional the performance would be even better. Besides measuring the success on community detection, the performance of the algorithm is also evaluated on the false detections: false communities and false neurons within good communities. Based on initial results, to minimise false communities we naively fixed a threshold for a minimum community size at $30$ neurons, considering everything smaller as an outcome of the random process used to generate the connections in the network. This limits the resolution of the algorithm: if there are real communities whose size is smaller than the threshold, they will not be detected. Since the average size of communities formed out of random depends on the size of the network, this part of the algorithm can be improved by using a threshold that is a function of $N$. Such a function can be derived by carrying out a systematic analysis on completely random networks, both theoretically and through simulations.

On the other hand, it is worth noting there is nothing that makes a false community different from a true community, except for the fact that the latter has gone through a learning process. A possible approach to solve this problem could be therefore to track the evolution of the connectivity whenever possible. After having detected the communities, looking back at the history of their connections could give important insights about which sets really experienced a learning process.

The algorithm we presented requires setting a number of thresholds, which makes the research highly customisable and it also allows different degrees of searching: for instance we can be interested in finding only highly significant bidirectional communities, if any. We can tune the thresholds as we like for a stricter search, which would require also less computational time. Once we have the outcome, we can then gradually relax the values of the thresholds for a broader search, if we need.

The full algorithm is the combination of two sub-algorithms executed one after the other. The first sub-algorithm alone already gives excellent results, especially for single community detection, with a massive reduction of the running time. Indeed, the second sub-algorithm is essential for dealing with overlaps and for resolving two communities that have been detected as a big one. Also, due to the statistical approach of the algorithm, it is possible to evaluate the level of noise (bidirectional pairs formed out of chance) and take it into account from the beginning of the procedure. This would allow to withdraw a consistent fraction of neurons before executing the two sub-algorithms and therefore to greatly reduce the number of neurons for the search. These aspects need to be further investigated, together with the possibility of introducing parallel computing, to improve on the computational time requirements.

\bibliographystyle{apalike}
\bibliography{biblio}

\section*{Acknowledgements}

We acknowledge support from the European Commission (FP7 Marie Curie Initial Training Network ``NAMASEN", grant n. 264872, http://cordis.europa.eu/fp7) and from the Engineering and Physical Sciences Research Council (grant n. EP/J019534/1). The funders had no role in study design, data collection and analysis, decision to publish, or preparation of the manuscript.

\section*{Author contributions statement}

U.E. E.V. conceived and designed the experiments. U.E. performed the experiments and analysed the data. U.E. E.V. wrote and revised the manuscript.

\section*{Additional information}

\textbf{Competing financial interests} The authors declare no competing financial interests.

\end{document}